\documentclass[manuscript,nonacm]{acmart} % 'nonacm' to remove copyright footer
\pdfoutput=1
\AtBeginDocument{%
  \providecommand\BibTeX{{%
    \normalfont B\kern-0.5em{\scshape i\kern-0.25em b}\kern-0.8em\TeX}}}

\acmConference[CoDaC '24]{Workshop on Correct Data Compression}{July 22,
  2024}{Montreal, CA}

\usepackage[justification=centering]{caption}
\usepackage{pgfplots}
\pgfplotsset{compat=1.18}
\usepackage{fancyvrb}

\usepackage{listings}
\usepackage{xcolor}

\definecolor{codegreen}{rgb}{0,0.6,0}
\definecolor{codegray}{rgb}{0.5,0.5,0.5}
\definecolor{codepurple}{rgb}{0.58,0,0.82}
\definecolor{backcolour}{rgb}{0.95,0.95,0.92}

\lstdefinestyle{mystyle}{
    backgroundcolor=\color{backcolour},   
    commentstyle=\color{codegreen},
    keywordstyle=\color{magenta},
    numberstyle=\tiny\color{codegray},
    stringstyle=\color{codepurple},
    basicstyle=\ttfamily\footnotesize,
    breakatwhitespace=false,         
    breaklines=true,                 
    captionpos=b,                    
    keepspaces=true,                 
    numbersep=5pt,                  
    showspaces=false,                
    showstringspaces=false,
    showtabs=false,                  
    tabsize=2
}

\begin{document}

% \title{Guaranteeing Error in Lossy Compression, Lessons Learned}
%\title{Lessons Learned on the Path to Guaranteed-Error Lossy Compression}
\title{Lessons Learned on the Path to Guaranteeing the Error Bound in Lossy Quantizers}

\author{Alex Fallin}
\affiliation{%
 \institution{Texas State University}
 \streetaddress{601 University Dr}
 \city{San Marcos}
 \state{Texas}
 \country{United States}}

\author{Martin Burtscher}
\affiliation{%
 \institution{Texas State University}
 \streetaddress{601 University Dr}
 \city{San Marcos}
 \state{Texas}
 \country{United States}}

\renewcommand{\shortauthors}{Fallin and Burtscher}

\begin{abstract}
Rapidly increasing data sizes in scientific computing are the driving force behind the need for lossy compression. The main drawback of lossy data compression is the introduction of error. This paper explains why many error-bounded compressors occasionally violate the error bound and presents the solutions we use in LC, a CPU/GPU compatible lossy compression framework, to guarantee the error bound for all supported types of quantizers. We show that our solutions maintain high compression ratios and cause no appreciable change in throughput.
\end{abstract}





\maketitle

\section{Introduction}
\label{sec:intro}
Many scientific instruments and simulations generate more data than can reasonably be handled, both in terms of throughput and in terms of total size~\cite{scientific_size}. There are two types of data compression, lossy and lossless, to alleviate these problems. Lossless compressors exactly reproduce the original data bit-for-bit. However, they are often not able to deliver the desired compression ratios. In contrast, lossy compression can yield much higher compression ratios, but with the caveat that the data is not exactly reproduced. High compression ratios without knowing the quality of the reconstructed data is not useful, thus lossy compression is often error-bounded. This means that the original value and the decompressed value do not deviate by more than a preset threshold. The three most frequently used error-bound types are point-wise absolute error (ABS), point-wise relative error (REL), and point-wise normalized absolute error (NOA). Bounding the error is important for scientific analysis %Unbounded error can achieve any arbitrary compression ratio at the cost of losing most of all of the information.
as it gives the users that decompress and analyze the data confidence that the information is reasonably preserved. Otherwise, they may not be able to draw valid conclusions from the decompressed data.

% When processing large datasets, high throughputs are generally desired and sometimes required. While GPUs are very effective at attaining high throughputs, their architecture differs greatly from that of CPUs. Moreover, domain scientists may not have a GPU in their workstations, so the ability to compress and decompress on either a CPU or GPU with bit-for-bit parity is an important consideration. Finally, the data producer and the data consumer are almost always two different systems, making consistency between systems crucial.
In today's heterogeneous HPC environments, scientific data is often generated and compressed on one type of device (i.e., a CPU or a GPU) but needs to be decompressed on a different type of device. For example, GPU-based compression may be critical for applications that produce data at a very high throughput, while CPU-based compression may be sufficient in other environments. Independently, the resulting data may be decompressed and analyzed by various users who may or may not have a GPU. Hence, cross-device compression and decompression is important, but it is rarely supported by today's state-of-the-art lossy compressors.

Beyond differences in hardware, issues with software, mainly surrounding the use of floating-point values and operations, can be the reason for failing to meet the error bound. Rounding issues, lack of associativity, compiler optimizations, and special values all have the potential to cause lack of parity or error-bound violations.

Allowing for error in compression begs two important questions. What factors affect the error bounding when reconstructing data? How do we guarantee matching results across different hardware (i.e., parity)? In this paper, we discuss our answers to these questions and describe the solutions we implemented in the guaranteed-error-bounded lossy quantizers of our open-source LC framework~\cite{lc_github}.

This paper makes the following main contributions.
\begin{itemize}
    \item It presents an evaluation of state-of-the-art lossy compressors on all possible single-precision floating-point values and many double-precision values showing that most of them violate the error bound on some values.
    \item It describes problems with floating-point arithmetic that can cause such violations.
    \item It discusses difference between CPUs and GPUs that cause them to produce different compressed data when running the same compression algorithm.
    \item It explains the code changes we had to make to guarantee the error bound and inter-device parity in all cases.
    \item It analyzes the impact of these changes on the compression ratio and throughput.
\end{itemize}

The rest of the paper is organized as follows. Section~\ref{sec:background} describes the common types of error-bounding. It further describes the problems we encountered, in terms of correctness, with floating-point arithmetic and parity between the CPU and GPU when implementing our compressor. Section~\ref{sec:appr} presents the code changes we made in our quantizers to address these floating-point arithmetic and parity issues. Section~\ref{sec:related} summarized related work on lossy compression and explains how various prior compressors run into correctness problems. Section~\ref{sec:methodology} describes our evaluation methodology. Section~\ref{sec:res} measures and discusses the impact of our solutions on the compression ratio and throughput. Section~\ref{sec:conclusion} concludes the paper with a summary.

\section{Background}
\label{sec:background}

%\subsection{Error-bound Types}

There are three point-wise error-control metrics that are commonly used in the literature: point-wise absolute (ABS), point-wise relative (REL), and point-wise normalized absolute (NOA). In this subsection, we define these error bounds and describe their uses~\cite{sdrbench}.

\subsection{Common Error-bound Types}

\subsubsection{Point-Wise Absolute Error (ABS)}
The point-wise {\em absolute} error of a data value is the {\em difference} between the original value of the data point and its reconstructed value~\cite{sz_abs_r2r}. The absolute error of a data value $x$ is defined as $e_{abs} = |x_{original} - x_{reconstructed}|$. To guarantee an absolute error bound of $\varepsilon$, each value in the reconstructed file must satisfy $e_{abs} \leq \varepsilon$. 
Hence, each reconstructed value must be in the range $x_{original} - \varepsilon \leq x_{reconstructed} \leq x_{original} + \varepsilon$.

%ABS error bounds are useful when the data is quite homogeneous in terms of magnitude or when the user does not have a particular interest in areas where values may be small relative to the error bound, that is, when the user cares mainly about the ``big picture''.

\subsubsection{Point-Wise Relative Error (REL)}
The point-wise {\em relative} error of a data value is the {\em ratio} between the difference of the original and its reconstructed value and the original value~\cite{sz_pwrel}. The relative error of a value $x$ is expressed as $e_{rel} = |\frac{x_{original} - x_{reconstructed}}{x_{original}}| = |1 - \frac{x_{reconstructed}}{x_{original}}|$. To guarantee a relative error bound of $\varepsilon$, every value in the reconstructed file must satisfy $e_{rel} \leq \varepsilon$.
In other words, each reconstructed value must have the same sign as the original value and be in the range $|x_{original}| / (1 + \varepsilon) \leq |x_{reconstructed}| \leq |x_{original}| \times (1 + \varepsilon)$. 

%REL error bounds are useful when the user wants to preserve a high level of detail in data areas where the values are close to zero but does not mind a higher absolute error in data areas where the absolute values are larger.

\subsubsection{Point-Wise Normalized Absolute Error (NOA)}
The point-wise {\em normalized absolute} error is the point-wise absolute error normalized by the value range $R = x_{max} - x_{min}$, that is, the range between the largest and the smallest value in the input.
The normalized absolute error of a data value $x$ is defined as $e_{noa} = |\frac{e_{abs}}{R}|$. To guarantee an error bound of $\varepsilon$, each value in the reconstructed file must satisfy $e_{noa} \leq \varepsilon$.
Hence, each reconstructed value must be in the  range $x_{original} - \varepsilon R \leq x_{reconstructed} \leq x_{original} + \varepsilon R$. Since NOA is a variant of and has the same issues as ABS, we do not separately evaluate it in this paper.

\subsection{Floating-point Arithmetic}
%Floating point math is inherently more complex than math outside of a computer.
%Because computers must represent all data in a finite number of bits, precision-sensitive applications must be extra careful to ensure that data is being correctly preserved. 

Floating-point values cannot precisely represent all numbers. Values that cannot be represented are rounded to a representable value. This behavior is important to note as rounding issues during data reconstruction are a common cause of error-bound violations. For example, ABS quantization is generally performed by multiplying the input value by the inverse of twice the error bound and rounding the result to the nearest (integer) bin number. This operation should be completely safe and yield a reconstructed value (i.e., the center of the value range represented by the bin) that differs from the original value by no more than the error bound. Small rounding issues, however, can cause an error-bound violation by placing a value that is very close to the border of one bin into the neighboring bin. Note that this is a problem even if the rounding error is much smaller than the error bound used for the quantization.
% should we mention the performance impact of denormals?

Infinity (INF), not-a-number (NaN), and denormal floating-point values pose additional challenges. INF and NaN values propagate when used in floating-point computations. Denormals are particularly susceptible to rounding issues as they are unable to retain the same precision as normal values. These special values, while problematic, must be preserved. For example, with an ABS error bound, normal and denormal values can be binned, but infinities and NaNs must be separately handled because, for example, $NaN \pm 0.001$ is still $NaN$ and $INF \pm 0.001$ is still $INF$. For a REL error bound, even denormals may require special handling.

\subsection{Result Parity}
In the process of developing our quantizers, we encountered many problems related to result parity between CPUs and GPUs. In this subsection, we describe these issues and give examples of code that causes them.

A fused multiply add (FMA) is a special machine instruction that performs both a multiplication and an addition without rounding the intermediate result, thus sometimes producing a different answer than a multiplication followed by a separate addition does. Since the FMA is faster, optimizing compilers try to use them when possible. For example, consider the following partial check of whether the ABS error bound has been exceeded (where \verb|eb2| is twice the error bound): \verb|bin * eb2 + eb < orig_value|. The left-hand-side expression may be compiled into an FMA depending on the many factors taken into account when optimizing the code. This optimization changes the rounding error as described above. What is more, different compilers make different optimization decisions as is the case for our CPU and GPU compilers, causing a disparity between the CPU compressed file and the GPU compressed file.
Additionally, as compilers evolve, code that does not currently yield FMA instructions may do so in the future.

Another major problem with supporting both CPUs and GPUs is the difference in libraries, and thus the results of some of the basic functions that would normally be expected to match. While developing the quantizers of LC, we encountered such a mismatch. REL uses the \verb|log()| function in the quantization and the \verb|pow()| function in the reconstruction step. Interestingly, these two functions do not produce the same result when passed the same argument on a CPU and a GPU. An actual example is the GPU producing a \verb|log()| result of $88.5$ when the CPU produces $88.4999...$. Whereas this mismatch seems small, it may result in one code choosing a different bin than the other, removing the guarantee of parity between the CPU and GPU. %In particular, this small mismatch can result in the GPU producing a different bin value than the CPU. Hence, low-level functions are potential sources of disparity.

\subsection{Edge Cases}

Beyond floating-point arithmetic and result parity, it is important to handle edge cases. For instance, we found that we cannot use \verb|std::abs()| in our quantizers. We originally used the check \verb|if ((std::abs(bin) >= maxbin) ...)| to determine if a bin number was valid. Since the range of twos-complement integers is $-2,147,483,648$ to $2,147,483,647$ (note the difference in the last digit), \verb|std::abs()| does not work for $-2,147,483,648$ as there is no corresponding positive value. While this is a 1-in-4-billion edge case, we encountered it on a real scientific input.

%The barriers to creating a lossy compressor that is able to correctly guarantee an error bound are individually small. The combination of these barriers, and the additional obstacles that exist for guaranteeing parity between devices, become quite an issue. It is clear as to why almost all of the related works are unable to fully guarantee error bound in all cases.

\section{Approach}
\label{sec:appr}

The previous section describes the three types of issues we had to contend with while developing the quantizers in LC~\cite{lc_github}, that is, the rounding of floating-point values, differences between CPUs and GPUs, and corner cases. In this section, we describe the solutions we implemented to create a CPU/GPU-compatible lossy compressor that provides a true error-bound guarantee and discuss the impacts these solutions have.

\subsection{Floating-point Arithmetic}
To address the rounding issues inherent to floating-point operations, we employ ``double-checking'' in the quantization step, meaning we immediately reconstruct each value and check whether it is within the error bound. %After quantization, all data is handled losslessly, but as mentioned above, we cannot trust the quantized bin to be reconstructed correct based on math rules alone.
To this end, we included the following lines of code, where \verb|bin| is the quantized bin number, \verb|eb| the error bound, \verb|eb2| twice the error bound, \verb|origval| the original value, and \verb|recon| the reconstructed value (we only show the relevant \verb|if| conditions):
%\begin{Verbatim}[fontsize=\small]
\begin{lstlisting}[language=C++]
const float recon = bin * eb2;
if (fabsf(origval - recon) > eb) ...
\end{lstlisting}
%\end{Verbatim}
By performing this check, we catch any floating-point issues that would cause the requested bound to be violated. If the condition is met, the value is preserved losslessly as we cannot quantize it within the error bound. For all three error bound types, we found that most of the input files contain at least one outlier that is caught by this test and preserved losslessly. We store these losslessly preserved outliers in-line with the bin numbers, which simplifies the program parallelization. This is in contrast to, for example, SZ3~\cite{sz3-1,sz3-2,sz3-3}, which does not commingle outliers and bin numbers but instead stores outliers in a separate list and uses bin number 0 to indicate an outlier.

%Infinity and NaN values are processed as follows.
We handle infinity by explicitly checking for it in our REL quantizer. In the ABS quantizer, the check is implicit; infinities are encoded losslessly because they cause checks that handle other error-bound issues to fail. Both quantizers explicitly check for and handle NaNs. Denormals are treated like normal values. % in the same if-statement as above, we check \verb|orig_f != orig_f| which always evaluates as false for NaN values.

\subsection{Result Parity}
The precision- and performance-increasing fused-multiply-add instructions can sometimes be avoided by tricking the compiler into thinking the intermediate value is used when it actually is not. This is not a reliable fix, however, as compiler improvements may be able to determine that the intermediate value goes unused. Therefore, we use the compiler flag \verb|-mno-fma| for $g$++ and the similar flag \verb|-fmad=false| for $nvcc$ to disable the use of FMAs. Note that these flags may potentially reduce the achievable precision. However, this is not a problem because of the aforementioned double checking. On the off chance that the reduced precision yields a wrong bin number, the corresponding value is simply encoded losslessly. This may lower the compression ratio, but it will not violate the error bound. Employing these flags, in combination with only using fully IEEE 754-compliant floating-point operations, results in code that produces the same compressed and decompressed values on CPUs and GPUs.

The differing \verb|log()| and \verb|pow()| functions were particularly challenging to fix. The solution we ultimately adopted is to write our own approximation functions. Our \verb|log2()| and \verb|pow2()| code for single-precision data is as follows:

\newpage

%\begin{figure}[htbp!]
\begin{lstlisting}[language=C++]
float log2approxf(const float orig_f) {
  const int mantissabits = 23;
  const int orig_i = *((int*)&orig_f); // extract bit pattern
  const int expo = (orig_i >> mantissabits) & 0xff; // isolate exponent
  const int frac_i = (127 << mantissabits) | (orig_i & ~(~0 << mantissabits)); // isolate fraction
  const float frac_f = *((float*)&frac_i); // convert fraction back to float
  const float log_f = frac_f + (expo - 128); // add de-biased exponent
  return log_f;
}
\end{lstlisting}
\vskip -2.5mm
\begin{lstlisting}[language=C++]
float pow2approxf(const float log_f) {
  const int mantissabits = 23;
  const float biased = log_f + 127; // re-bias exponent
  const int expo = biased; // get exponent
  const float frac_f = biased - (expo - 1); // recreate fraction
  const int frac_i = *((int*)&frac_f); // extract fraction
  const int exp_i = (expo << mantissabits) | (frac_i & ~(~0 << mantissabits)); // combine exp & frac
  const float recon_f = *((float*)&exp_i); // convert back to float
  return recon_f;
}
\end{lstlisting}
%\caption{}
%\label{lst:logpow}
%\end{figure}

These approximations guarantee matching solutions between the CPU and GPU because every operation within them is fully IEEE 754-compliant or an integer operation. As shown in Section~\ref{sec:res}, this solution hurts the compression ratio a little because the approximation is not particularly accurate. As before, it does not affect correctness because results that exceed the error bound are discarded and the corresponding values losslessly encoded.

\subsection{Edge Case}

We handle the problem with \verb|std::abs()| by breaking the single \verb|(std::abs(bin) >= maxbin)| check into two separate checks, namely \verb$((bin >= maxbin) || (bin <= -maxbin))$. This fixes the edge case but requires an additional check.

\section{Related Work}
\label{sec:related}

%> Let’s just discuss the compressors listed in the related work of the PFPL paper but change the discussion to focus on what types of error they support and whether they guarantee the bounds (and, if so, how).

As lossy compression is a widely researched domain, this section focuses on the lossy floating-point compressors we evaluate in Section~\ref{sec:res}. Table~\ref{tab:related_table} summarizes these compressors and their support for the widely-used error-bound types. A `{\color[HTML]{70AD47}\textbf{\checkmark}}' indicates that the compressor supports that error-bound type whereas a `{\color[HTML]{FFA500}$\circ$}' shows that it does not.

\begin{table*}[!hbtp]
\centering
\caption{All tested compressors and the error-bound types they support at a glance (ordered by initial release date)}
\label{tab:related_table}
\resizebox{.4\textwidth}{!}{
\begin{tabular}{lcccccccccccc}
\centering
 & \multicolumn{1}{c}{} & \multicolumn{1}{c}{} & \multicolumn{1}{c}{} & \multicolumn{1}{c}{Guaranteed}\\
Compressor & \multicolumn{1}{c}{ABS} & \multicolumn{1}{c}{REL} & \multicolumn{1}{c}{NOA} & \multicolumn{1}{c}{error bound}\\ \hline
ZFP & {\color[HTML]{70AD47} \textbf{\checkmark}} & {\color[HTML]{FFA500}$\circ$} & {\color[HTML]{FFA500}$\circ$} & {\color[HTML]{FFA500}$\circ$} \\ \hline
SZ2 & {\color[HTML]{70AD47} \textbf{\checkmark}} & {\color[HTML]{70AD47} \textbf{\checkmark}} & {\color[HTML]{70AD47} \textbf{\checkmark}} & {\color[HTML]{FFA500}$\circ$} \\ \hline
SZ3 & {\color[HTML]{70AD47} \textbf{\checkmark}} & {\color[HTML]{FFA500}$\circ$} & {\color[HTML]{70AD47} \textbf{\checkmark}} & {\color[HTML]{70AD47} \textbf{\checkmark}} \\ \hline
MGARD-X & {\color[HTML]{70AD47} \textbf{\checkmark}} & {\color[HTML]{FFA500}$\circ$} & {\color[HTML]{70AD47} \textbf{\checkmark}} & {\color[HTML]{FFA500}$\circ$} \\ \hline
SPERR & {\color[HTML]{70AD47} \textbf{\checkmark}} & {\color[HTML]{FFA500}$\circ$} & {\color[HTML]{FFA500}$\circ$} & {\color[HTML]{FFA500}$\circ$} \\ \hline
FZ-GPU & {\color[HTML]{FFA500}$\circ$} & {\color[HTML]{FFA500}$\circ$} & {\color[HTML]{70AD47} \textbf{\checkmark}} & {\color[HTML]{FFA500}$\circ$} \\ \hline
cuSZp & {\color[HTML]{70AD47} \textbf{\checkmark}} & {\color[HTML]{FFA500}$\circ$} & {\color[HTML]{70AD47} \textbf{\checkmark}} & {\color[HTML]{FFA500}$\circ$} \\ \hline
\textbf{LC} & {\color[HTML]{70AD47} \textbf{\checkmark}} & {\color[HTML]{70AD47} \textbf{\checkmark}} & {\color[HTML]{70AD47} \textbf{\checkmark}} & {\color[HTML]{70AD47} \textbf{\checkmark}} \\ \hline
\end{tabular}}
\end{table*} % Feature Table

There are four main versions of SZ. They all use prediction in their compression pipeline. SZ2~\cite{sz2} employs Lorenzo prediction~\cite{lorenzo} and linear regression followed by quantization and lossless compression. SZ3~\cite{sz3-1,sz3-2,sz3-3} is an improvement that typically produces better compression ratios with similar throughput. It adds preprocessing before the prediction and entropy coding to the lossless compression stage. SZ2 and SZ3 are both CPU-only compressors. As discussed in Section~\ref{sec:background}, outliers are likely to occur. While LC leaves these outliers in-line, SZ3 uses the `0' bin as a reserved value for outliers, which are grouped outside of the quantized portion. cuSZ~\cite{cusz-1,cusz-2} is a CUDA implementation that employs a different, more GPU-friendly algorithm. It performs Lorenzo prediction and quantization followed by multi-byte Huffman coding. FZ-GPU~\cite{fzgpu} is a specialized version of cuSZ that fuses multiple kernels together for better throughput. cuSZp~\cite{cuszp} splits the data into blocks and then quantizes and predicts the values in all nonzero blocks. Next, it losslessly compresses the result. Similar to LC, SZ2 and SZ3 control the error by reconstructing the value in the compression stage. They tighten the error bound for values that would otherwise exceed the error bound. FZ-GPU and cuSZp both quantize in the same way that LC does. Unlike LC, however, they do not double-check whether the quantization is within the requested error bound. All versions of SZ support ABS error-bounding, but only SZ2 supports REL error-bounding. They also all support single-precision data, and only FZ-GPU does not support double-precision values.

ZFP~\cite{zfp,zfpnew} is a widely used compression tool that is based on a custom decorrelating transform. It is specifically designed for in-memory array compression and supports on-the-fly random-access decompression. ZFP splits the input into blocks, converts each value into an integer, performs the aforementioned decorrelation, reorders the data, and converts the values to negabinary representation. Then, it shuffles the bits and losslessly compresses them. ZFP controls the error during the transformation into an integer. The theorem used to support error guarantees assumes infinite precision. Due to this assumption, ZFP is susceptible to floating-point arithmetic errors in some cases. It supports the ABS error-bound and both single- and double-precision data.

MGARD~\cite{mgard,mgardx} is the only other compressor we found that also supports compatible compression and decompression between CPUs and GPUs. This compressor uses multi-grid hierarchical data refactoring to decompose the data into coefficients with correction factors for reconstruction. The error is controlled during decompression by selectively loading the correct hierarchy of decomposed data based on the requested error bound. It supports the ABS error bound and both single- and double-precision data.

SPERR~\cite{sperr}, which is an evolution of SPECK~\cite{speck}, uses advanced wavelet transforms that are applied recursively to the input. SPERR detects outliers that do not meet the error bound and stores correction factors for those values. This correction appears to be susceptible to floating-point arithmetic errors, especially as outliers are refined in further steps. SPERR supports ABS error-bounding and both types of floating-point data.

% Table~\ref{tab:feature_table} shows the compressors we compare to along with our PFPL approach. The top row lists features relevant to this paper; each remaining row corresponds to one compressor. A `{\color[HTML]{70AD47}\textbf{\checkmark}}' indicates that the corresponding compressor supports the feature whereas an `{\color[HTML]{FF0000}$\times$}' indicates that we found the compressor to not support it. `ABS' and `REL' denote the error-bound type. `Float' and `Double' indicate support for 32-bit single-precision and 64-bit double-precision data. `CPU' and `GPU' display support for compression and decompression on the respective device. This spread of compressors represents many approaches to guaranteeing error bound and, as seen in Section~\ref{sec:res}, is a good example of how complex the issue is.

\section{Experimental Methodology}
\label{sec:methodology}

We evaluated the compressors described in %Table~\ref{tab:feature_table} 
Section~\ref{sec:related} on a system based on an AMD Ryzen Threadripper 2950X CPU with 16 cores. Hyperthreading is enabled, that is, the 16 cores can simultaneously run 32 threads. The main memory has a capacity of 64 GB. The operating system is Fedora 37. The GPU is an NVIDIA RTX 4090 (Ada Lovelace architecture) with 16,384 processing elements distributed over 128 multiprocessors. Its global memory has a capacity of 24 GB. The GPU driver version is 525.85.05. The GPU codes were compiled with {\em nvcc} version 12.0.140 using the ``-O3 -arch=sm\_89'' flags. Unless otherwise specified by the build process, we compiled the C++ codes using the ``-O3 -march=native'' flags.

When evaluating throughput, we measured the execution time of the compression and decompression functions, excluding any time spent reading the input file, verifying the results, and, for the GPU codes, transferring data to and from the GPU.
We run each experiment 9 times and collect the compression ratio, median compression throughput, and median decompression throughput. 

\begin{table}[hbtp]
    \begin{center}
        \caption{Information about the used input datasets}
        \vskip -3mm
        \label{tab:inputs}
        \resizebox{0.65\textwidth}{!}{
        \begin{tabular} { |l|l|l|c|c| }\hline
            \textbf{Name} &\textbf{Description}  &\textbf{Format} &\textbf{Files} &\textbf{Dimensions}\\\hline
                CESM-ATM &Climate &Single &33 &26 $\times$ 1800 $\times$ 3600\\\hline
                EXAALT Copper &Molecular Dynamics &Single &6 &Various 2D\\\hline
                HACC &Cosmology &Single &6 &280,953,867\\\hline
                Hurricane ISABEL &Weather Simulation &Single &13 &100 $\times$ 500 $\times$ 500\\\hline
                NYX &Cosmology &Single &6 &512 $\times$ 512 $\times$ 512\\\hline
                QMCPACK &Quantum Monte Carlo &Single &2 &33,120 $\times$ 69 $\times$ 69\\\hline
                SCALE &Climate &Single &12 &98 $\times$ 1200 $\times$ 1200\\\hline
        \end{tabular}}
    \end{center} 
\end{table}

We used the 7 single-precision suites shown in Table~\ref{tab:inputs} as inputs. They stem from the SDRBench repository~\cite{sdrbench_url,sdrbench}, which hosts scientific datasets from different domains. For the throughput evaluation, we use only one file from each input set because the performance of our compressor does not change significantly between individual inputs within a suite. For the compression-ratio evaluation, we use all the inputs and report the geometric mean within each suite. % , which hosts scientific datasets for compression evaluation. The table lists the suite's name, a short description, floating-point format, number of datasets, and input dimensions.

% To keep the number of inputs reasonable we excluded some datasets.
% We only use the 3D CESM-ATM inputs as they contain similar data to the other CESM-ATM inputs and 3D is a common dimension in scientific applications. We use only the EXAALT Copper dataset because it is in the middle in terms of size for the EXAALT sets. We use the raw (i.e., not cleared) data from the Hurricane ISABEL set. Additionally, we exclude SDRBench datasets that are either too large or in a proprietary format.

Additionally, we generated sets of single- and double-precision inputs that cover a wide range of values, including positive and negative infinity (INF), not-a-number (NaN), and denormal values, which sometimes cause issues in floating-point compressors. As mentioned, we only test ABS and REL error bounds as NOA is similar to ABS.

%The problematic value evaluation is presented in the form of a table where the problematic value functionality of each compressor is shown. A `{\color[HTML]{70AD47}\textbf{\checkmark}}' is used to indicate the compressor successfully handles that value. A `{\color[HTML]{FFA500}$\circ$}' shows that the compressor does not guarantee the error bound, but also does not crash. Last, a `{\color[HTML]{FF0000}$\times$}' denotes a crash when supplied that value.

We report the throughput and compression results in bar charts where the bars are the metric in question normalized to the non-correctness-guaranteed metric. We use the REL quantizer for the \verb|pow()| and \verb|log()| comparisons because only REL requires these functions. We use the ABS quantizer to evaluate the rounding-error protection.

\section{Results}
\label{sec:res}

%In this section, we show the compression ratio and throughput implications of some of the solutions discussed in Section~\ref{sec:appr}.
Table~\ref{tab:funny_values} summarizes which kinds of values each tested compressor can handle. The results are for ABS only, with the exception of SZ2 and LC, which support both REL and ABS. A `{\color[HTML]{70AD47}\textbf{\checkmark}}' indicates that the compressor successfully handles this kind of value. A `{\color[HTML]{FFA500}$\circ$}' shows that the compressor does not guarantee the error bound but also does not crash. Finally, a `{\color[HTML]{FF0000}$\times$}' denotes a crash when supplied that kind of value.

\begin{table}[htbp]
    \begin{center}
        \caption{Values that meet the error bound. `{\color[HTML]{70AD47}\textbf{\checkmark}}' indicates no issues, `{\color[HTML]{FFA500}$\circ$}' indicates error-bound violations, and `{\color[HTML]{FF0000}$\times$}' indicates a crash.}
        \vskip -3mm
        \label{tab:funny_values}
        \resizebox{0.6\textwidth}{!}{
        \begin{tabular}{l|c|ccc|ccc|}
        \cline{2-8}
         & \multicolumn{4}{c|}{Single} & \multicolumn{3}{c|}{Double} \\ \hline
        \multicolumn{1}{|l|}{Compressor} & \multicolumn{1}{l|}{Normal} & \multicolumn{1}{l|}{INF} & \multicolumn{1}{l|}{NaN} & \multicolumn{1}{l|}{Denormal} & \multicolumn{1}{l|}{INF} & \multicolumn{1}{l|}{NaN} & \multicolumn{1}{l|}{Denormal} \\ \hline
        \multicolumn{1}{|l|}{ZFP} & {\color[HTML]{FFA500}$\circ$} & \multicolumn{1}{c|}{{\color[HTML]{FFA500}$\circ$}} & \multicolumn{1}{c|}{{\color[HTML]{FFA500}$\circ$}} & {\color[HTML]{70AD47}\textbf{\checkmark}} & \multicolumn{1}{c|}{{\color[HTML]{FFA500}$\circ$}} & \multicolumn{1}{c|}{{\color[HTML]{FFA500}$\circ$}} & {\color[HTML]{70AD47}\textbf{\checkmark}} \\ \hline
        \multicolumn{1}{|l|}{SZ2} & {\color[HTML]{FFA500}$\circ$} & \multicolumn{1}{c|}{{\color[HTML]{70AD47}\textbf{\checkmark}}} & \multicolumn{1}{c|}{{\color[HTML]{70AD47}\textbf{\checkmark}}} & {\color[HTML]{FFA500}$\circ$} & \multicolumn{1}{c|}{{\color[HTML]{70AD47}\textbf{\checkmark}}} & \multicolumn{1}{c|}{{\color[HTML]{70AD47}\textbf{\checkmark}}} & {\color[HTML]{FFA500}$\circ$} \\ \hline
        \multicolumn{1}{|l|}{SZ3} & {\color[HTML]{70AD47}\textbf{\checkmark}} & \multicolumn{1}{c|}{{\color[HTML]{70AD47}\textbf{\checkmark}}} & \multicolumn{1}{c|}{{\color[HTML]{70AD47}\textbf{\checkmark}}} & {\color[HTML]{70AD47}\textbf{\checkmark}} & \multicolumn{1}{c|}{{\color[HTML]{70AD47}\textbf{\checkmark}}} & \multicolumn{1}{c|}{{\color[HTML]{70AD47}\textbf{\checkmark}}} & {\color[HTML]{70AD47}\textbf{\checkmark}} \\ \hline
        \multicolumn{1}{|l|}{MGARD-X} & {\color[HTML]{FFA500}$\circ$} & \multicolumn{1}{c|}{{\color[HTML]{70AD47}\textbf{\checkmark}}} & \multicolumn{1}{c|}{{\color[HTML]{70AD47}\textbf{\checkmark}}} & {\color[HTML]{70AD47}\textbf{\checkmark}} & \multicolumn{1}{c|}{{\color[HTML]{70AD47}\textbf{\checkmark}}} & \multicolumn{1}{c|}{{\color[HTML]{70AD47}\textbf{\checkmark}}} & {\color[HTML]{70AD47}\textbf{\checkmark}} \\ \hline
        \multicolumn{1}{|l|}{SPERR} & {\color[HTML]{FFA500}$\circ$} & \multicolumn{1}{c|}{{\color[HTML]{FF0000}$\times$}} & \multicolumn{1}{c|}{{\color[HTML]{FF0000}$\times$}} & {\color[HTML]{70AD47}\textbf{\checkmark}} & \multicolumn{1}{c|}{{\color[HTML]{FF0000}$\times$}} & \multicolumn{1}{c|}{{\color[HTML]{FF0000}$\times$}} & {\color[HTML]{70AD47}\textbf{\checkmark}} \\ \hline
        \multicolumn{1}{|l|}{FZ-GPU} & {\color[HTML]{FFA500}$\circ$} & \multicolumn{1}{c|}{{\color[HTML]{70AD47}\textbf{\checkmark}}} & \multicolumn{1}{c|}{{\color[HTML]{70AD47}\textbf{\checkmark}}} & {\color[HTML]{70AD47}\textbf{\checkmark}} & \multicolumn{1}{c|}{n/a} & \multicolumn{1}{c|}{n/a} & n/a \\ \hline
        \multicolumn{1}{|l|}{cuSZp} & {\color[HTML]{FFA500}$\circ$} & \multicolumn{1}{c|}{{\color[HTML]{FF0000}$\times$}} & \multicolumn{1}{c|}{{\color[HTML]{70AD47}\textbf{\checkmark}}} & {\color[HTML]{70AD47}\textbf{\checkmark}} & \multicolumn{1}{c|}{{\color[HTML]{FF0000}$\times$}} & \multicolumn{1}{c|}{{\color[HTML]{FF0000}$\times$}} & {\color[HTML]{70AD47}\textbf{\checkmark}} \\ \hline
        \multicolumn{1}{|l|}{LC} & {\color[HTML]{70AD47}\textbf{\checkmark}} & \multicolumn{1}{c|}{{\color[HTML]{70AD47}\textbf{\checkmark}}} & \multicolumn{1}{c|}{{\color[HTML]{70AD47}\textbf{\checkmark}}} & {\color[HTML]{70AD47}\textbf{\checkmark}} & \multicolumn{1}{c|}{{\color[HTML]{70AD47}\textbf{\checkmark}}} & \multicolumn{1}{c|}{{\color[HTML]{70AD47}\textbf{\checkmark}}} & {\color[HTML]{70AD47}\textbf{\checkmark}} \\ \hline
        \end{tabular}
        }
    \end{center}
\end{table} % Funny value table

All tested compressors can handle normal values, though most of them do not guarantee the error bound. These error-bound violations are likely due to the rounding issues discussed in Section~\ref{sec:background}. Every compressor except for SZ2 correctly handles and error-bounds denormal values. The reasons SZ2 does not properly handle these values is due to it being the only compressor (aside from LC) that supports REL. When a small denormal value is bound using REL, it is highly susceptible to rounding issues. Three compressors have problems with INF and NaN, on which SPERR and cuSZp occasionally crash and ZFP is unable to guarantee the error bound.

The table highlights that even state-of-the-art compressors have problems with some values. By implementing the fixes discussed in Section~\ref{sec:appr}, LC is able to avoid crashing or violating the error bound on any of these values. In fact, we exhaustively tested it  on all roughly 4 billion possible 32-bit floating-point values with several error bounds to ensure that it handles all values correctly.

\begin{figure}[!htbp]
    \centering
        \begin{tikzpicture}
        \begin{axis}[
                ybar,
                bar width=.75cm,
                width=\textwidth,
                height=.22\textwidth,
                legend style={at={(0.5,1)},
                    anchor=north,legend columns=-1},
                symbolic x coords={CESM, EXAALT, HACC, NYX, QMCPACK, SCALE, ISABEL},
                xtick=data,
                nodes near coords,
                nodes near coords align={vertical},
                ymin=0,ymax=1.25,
                ylabel={\% of unmodified CR},
            ]
            \addplot table[x=interval,y=new]{\crdata};
        \end{axis}
    \end{tikzpicture}
    \vskip -3mm
    \caption{Compression ratios of the parity-ensured $pow()$ and $log()$ REL compressor normalized to the non-ensured ratios.}
    \label{fig:pow_log_cr}
\end{figure}
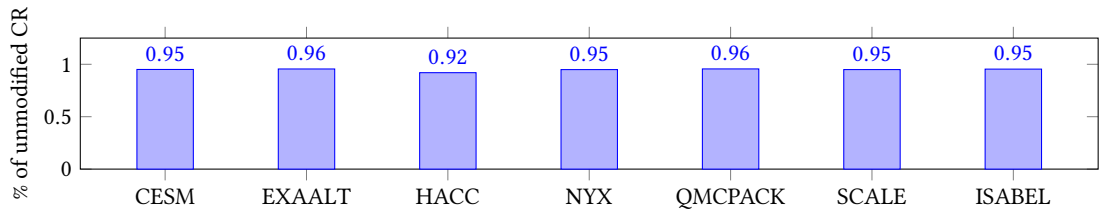

\begin{figure}[!htbp]
    \centering
        \begin{tikzpicture}
        \begin{axis}[
                ybar,
                bar width=.5cm,
                width=\textwidth,
                height=.22\textwidth,
                legend style={at={(0.5,1)},
                    anchor=north,legend columns=-1},
                symbolic x coords={CESM, EXAALT, HACC, NYX, QMCPACK, SCALE, ISABEL},
                xtick=data,
                nodes near coords,
                nodes near coords align={vertical},
                ymin=0,ymax=1.25,
                ylabel={\% of unmodified TP},
            ]
            \addplot table[x=interval,y=comp]{\pltpdata};
            \addplot table[x=interval,y=decomp]{\pltpdata};
        \end{axis}
    \end{tikzpicture}
    \vskip -3mm
    \caption{GPU throughput of the parity-ensured $pow()$ and $log()$ REL compressor normalized to the non-ensured throughputs. Blue bars are compression, red bars are decompression.}
    \label{fig:pow_log_tp}
\end{figure}
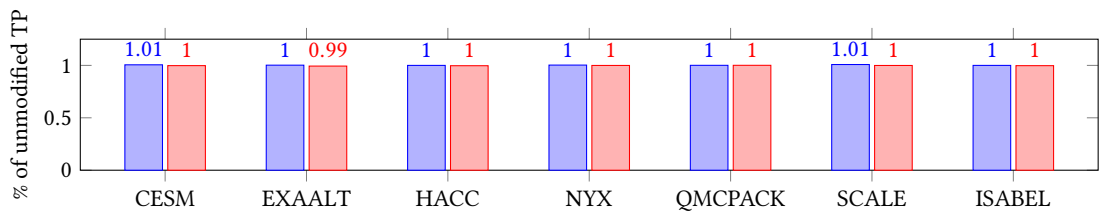

Figure~\ref{fig:pow_log_cr} and Table~\ref{tab:pow_log_cr_tab} show the compression ratio effects of our \verb|pow()| and \verb|log()| replacements in the REL quantizer for an error bound of 1E-3, and Figure~\ref{fig:pow_log_tp} and Tables~\ref{tab:pow_log_ctp_tab} and~\ref{tab:pow_log_dtp_tab} show the throughput effects. In the figures, the bar height indicates the performance relative to using the library versions of \verb|pow()| and \verb|log()| (higher is better). Each bar represents the geometric mean over all files in one input dataset for the compression ratios. %The variance of compression ratios within some of these datasets can be high because some of the files are very compressible while others within that same dataset are not. This is variance is not unique to LC as it is from the properties of the datasets.
For the throughput results, each bar represents the median GPU throughput for the representative file from that suite. %The throughput variance within a dataset is lower as each file from the same dataset is the same size and much of the work is memory-bound. It is slightly affected by the compression ratio due to smaller output sizes yielding better throughput performance.

When switching to our less-accurate but parity-protected approximation functions, the compression ratio is affected negatively, as expected. This drop in compression ratio is due to more values being stored losslessly as they are unable to be quantized within the error bound. Losslessly stored float values are harder to compress in the later stages of LC. While the compression loss is significant at 5.2\% on average, without it the compressor would be unable to produce the same result on both CPUs and GPUs. Note that this only affects the REL quantizer as the ABS and NOA quantizers do not use these approximation functions.

\pgfplotstableread[row sep=\\,col sep=&]{
    interval & orig & new \\
    CESM    & 1 & 0.95065332\\
    EXAALT  & 1 & 0.955298446\\
    HACC    & 1 & 0.919947177\\
    NYX     & 1 & 0.949388055\\
    QMCPACK & 1 & 0.955789236\\
    SCALE   & 1 & 0.949591164\\
    ISABEL  & 1 & 0.953951416\\
}\crdata % Data for CR table   
 % Pow and Log CR comparison

\begin{table}[!htbp]
\caption{Compression ratios with and w/o the parity-ensured $pow()$ and $log()$ functions in the REL compressor.}
\label{tab:pow_log_cr_tab}
\begin{tabular}{l|r|r|r|r|r|r|r|}
\cline{2-8}
 & CESM & EXAALT & HACC & NYX & QMCPACK & SCALE & ISABEL \\ \hline
\multicolumn{1}{|l|}{Original Functions} & 7.2 & 3.8 & 5.1 & 4.0 & 2.6 & 7.4 & 5.2 \\ \hline
\multicolumn{1}{|l|}{Replaced Functions} & 6.8 & 3.6 & 4.7 & 3.8 & 2.5 & 7.1 & 4.9 \\ \hline
\end{tabular}
\end{table} % Pow and Log CR table

\pgfplotstableread[row sep=\\,col sep=&]{
    interval & comp & decomp \\
    CESM    & 1.005567452 & 0.997836775\\
    EXAALT  & 1.002169842 & 0.993476401\\
    HACC    & 1.000000000 & 0.996960842\\
    NYX     & 1.002958580 & 0.999508479\\
    QMCPACK & 1.000671892 & 1.000828157\\
    SCALE   & 1.008300908 & 0.999289100\\
    ISABEL  & 1.000000000 & 0.997458704\\
}\pltpdata % Data for Pow and Log TP table
 % Pow and Log TP comparison

\begin{table}[!htbp]
\caption{GPU compression throughput in GB/s with and w/o the parity-ensured $pow()$ and $log()$ functions in the REL compressor.}
\label{tab:pow_log_ctp_tab}
\begin{tabular}{l|r|r|r|r|r|r|r|}
\cline{2-8}
 & CESM & EXAALT & HACC & NYX & QMCPACK & SCALE & ISABEL \\ \hline
\multicolumn{1}{|l|}{Original Functions} & 143.5 & 146.5 & 143.0 & 144.0 & 141.2 & 145.2 & 138.7 \\ \hline
\multicolumn{1}{|l|}{Replaced Functions} & 144.3 & 146.8 & 143.0 & 144.4 & 141.3 & 146.4 & 138.7 \\ \hline
\end{tabular}
\end{table} % Pow and Log CTP table

\begin{table}[!htbp]
\caption{GPU decompression throughput in GB/s with and w/o the parity-ensured $pow()$ and $log()$ functions in the REL compressor.}
\label{tab:pow_log_dtp_tab}
\begin{tabular}{l|r|r|r|r|r|r|r|}
\cline{2-8}
 & CESM & EXAALT & HACC & NYX & QMCPACK & SCALE & ISABEL \\ \hline
\multicolumn{1}{|l|}{Original Functions} & 132.8 & 134.7 & 131.8 & 132.0 & 130.5 & 133.9 & 127.4 \\ \hline
\multicolumn{1}{|l|}{Replaced Functions} & 132.5 & 133.8 & 131.4 & 131.9 & 130.6 & 133.8 & 127.1 \\ \hline
\end{tabular}
\end{table} % Pow and Log DTP table

Interestingly, the throughput of LC remains within $\pm 1\%$ when switching to our approximate functions. There are several reasons for why there is essentially no change in throughput. First, these functions only represent a small fraction of the overall execution time. Second, compression and decompression are memory-bound operations that may hide some of the computation latency. Third, the native \verb|pow()| and \verb|log()| are also quite slow and probably not much faster than our approximate functions, which exclusively use fast operations.
%Even if they were significantly slower, the affect on throughput as a whole would still be relatively small because memory access tends to control runtime for our code.

\begin{figure}[!htbp]
    \centering
        \begin{tikzpicture}
        \begin{axis}[
                ybar,
                bar width=.75cm,
                width=\textwidth,
                height=.22\textwidth,
                legend style={at={(0.5,1)},
                    anchor=north,legend columns=-1},
                symbolic x coords={CESM, EXAALT, HACC, NYX, QMCPACK, SCALE, ISABEL},
                xtick=data,
                nodes near coords,
                nodes near coords align={vertical},
                ymin=0,ymax=1.25,
                ylabel={\% of unprotected TP},
            ]
            \addplot table[x=interval,y=comp]{\dctpdata};
        \end{axis}
    \end{tikzpicture}
    \vskip -3mm
    \caption{GPU compression throughput of the rounding-error protected ABS compressor normalized to the non-protected throughput.}
    \label{fig:checks_tp}
\end{figure}
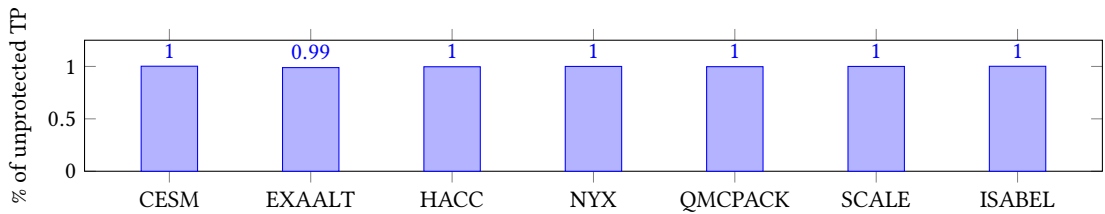

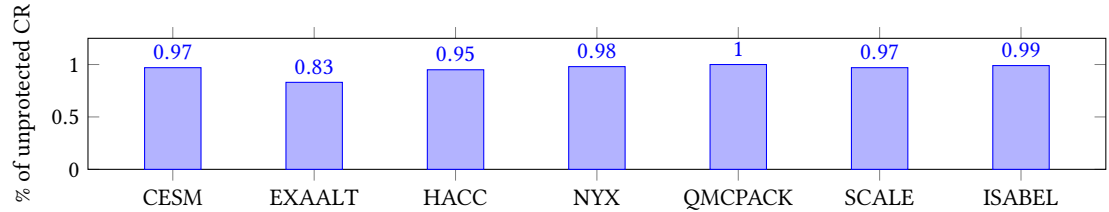
\begin{figure}[!htbp]
    \centering
        \begin{tikzpicture}
        \begin{axis}[
                ybar,
                bar width=.75cm,
                width=\textwidth,
                height=.22\textwidth,
                legend style={at={(0.5,1)},
                    anchor=north,legend columns=-1},
                symbolic x coords={CESM, EXAALT, HACC, NYX, QMCPACK, SCALE, ISABEL},
                xtick=data,
                nodes near coords,
                nodes near coords align={vertical},
                ymin=0,ymax=1.25,
                ylabel={\% of unprotected CR},
            ]
            \addplot table[x=interval,y=comp]{\dctpdata};
        \end{axis}
    \end{tikzpicture}
    \vskip -3mm
    \caption{GPU compression ratio of the rounding-error protected ABS compressor normalized to the non-protected compression ratio.}
    \label{fig:checks_abs_cr}
\end{figure}

Figure~\ref{fig:checks_tp} and Table~\ref{tab:checks_tp_tab} show the GPU throughput changes due to the rounding-error protection in our code using the ABS quantizer with an error bound of 1E-3 while Figure~\ref{fig:checks_abs_cr} and Table~\ref{tab:checks_cr_tab} show the compression ratio changes. For the bar charts, the height of the bar is normalized to the performance of the code without rounding-error protection. %We do not show compression ratios as rounding errors are a correctness issue and not a parity issue. Additionally, 
We do not show decompression results as the ``double checking'' is not present in the decompressor.

The addition of the extra checks to prevent an error-bound violation does not significantly affect throughput. The reasoning is likely the same as above. These checks represent very little of the total runtime and may be hidden under the memory-access latency. A bigger difference is observed in compression ratio. The version of the compressor with the double checking yields ratios that are about 5\% worse than the compressor that does not include the check.

Table~\ref{tab:rounding_errors} sheds light on the reason for this loss, where the most pronounced decrease in compression ratio corresponds to the highest percentage of values incurring rounding errors that must be mitigated. The loss in compression ratio is most pronounced in the EXAALT input set, which includes a file where 11.2\% of the values fail the verification. Nevertheless, every dataset compresses well, even though they all incur some rounding errors. They still compress well because all values, even the ones found to be non-quantizable due to a rounding error, are compressed losslessly. This helps mitigate the effect of non-quantizable values on compression ratio. Overall, this small loss in compression ratio is the cost of guaranteeing the error bound when floating-point arithmetic is involved.

\pgfplotstableread[row sep=\\,col sep=&]{
    interval & comp \\
    CESM    & 1.002083816 \\
    EXAALT  & 0.988815479 \\
    HACC    & 0.997033375 \\
    NYX     & 1.000000000 \\
    QMCPACK & 0.997721577 \\
    SCALE   & 1.000000000 \\
    ISABEL  & 1.001418440 \\
}\dctpdata % Data for rounding error TP table
 % Double-check TP comparison

\begin{table}[!htbp]
\caption{GPU compression throughput in GB/s of the rounding-error protected ABS compressor vs.~the non-protected compressor.}
\label{tab:checks_tp_tab}
\begin{tabular}{l|r|r|r|r|r|r|r|}
\cline{2-8}
 & CESM & EXAALT & HACC & NYX & QMCPACK & SCALE & ISABEL \\ \hline
\multicolumn{1}{|l|}{Protected} & 156.0 & 145.4 & 138.9 & 144.7 & 143.7 & 190.6 & 141.8 \\ \hline
\multicolumn{1}{|l|}{Unprotected} & 155.7 & 147.1 & 139.3 & 144.7 & 144.0 & 190.6 & 141.6 \\ \hline
\end{tabular}
\end{table}

\pgfplotstableread[row sep=\\,col sep=&]{
    interval & comp \\
    CESM    & .97 \\
    EXAALT  & .83 \\
    HACC    & .95 \\
    NYX     & .98 \\
    QMCPACK & 1 \\
    SCALE   & .97 \\
    ISABEL  & .99 \\
}\dctpdata % Data for rounding error TP table
 % Double-check CR comparison ABS

\begin{table}[!htbp]
\caption{Compression ratio of the rounding-error protected ABS compressor vs.~the non-protected compressor.}
\label{tab:checks_cr_tab}
\begin{tabular}{l|r|r|r|r|r|r|r|}
\cline{2-8}
 & CESM & EXAALT & HACC & NYX & QMCPACK & SCALE & ISABEL \\ \hline
\multicolumn{1}{|l|}{Protected} & 122.0 & 3.3 & 2.3 & 1.9 & 4.3 & 81.1 & 140.8 \\ \hline
\multicolumn{1}{|l|}{Unprotected} & 126.1 & 4.0 & 2.4 & 1.9 & 4.3 & 83.8 & 142.4 \\ \hline
\end{tabular}
\end{table}

\begin{table}[]
\caption{Percentage of the input values affected by rounding errors in the ABS quantizer.}
\label{tab:rounding_errors}
\begin{tabular}{c|r|r|}
\cline{2-3}
\multicolumn{1}{l|}{} & \multicolumn{1}{c|}{Average} & \multicolumn{1}{c|}{Maximum} \\ \hline
\multicolumn{1}{|c|}{CESM} & 0.12\% & 1.68\% \\ \hline
\multicolumn{1}{|c|}{EXAALT} & 3.41\% & 11.16\% \\ \hline
\multicolumn{1}{|c|}{HACC} & 0.25\% & 0.40\% \\ \hline
\multicolumn{1}{|c|}{NYX} & 0.89\% & 5.29\% \\ \hline
\multicolumn{1}{|c|}{QMCPACK} & 0.00\% & 0.00\% \\ \hline
\multicolumn{1}{|c|}{SCALE} & 0.16\% & 1.38\% \\ \hline
\multicolumn{1}{|c|}{ISABEL} & 0.05\% & 0.63\% \\ \hline
\end{tabular}
\end{table}

In summary, the solutions to the correctness problems we discovered while developing LC do not adversely affect the throughput but do lower the compression ratio noticeably. However, they guarantee the error bound for both ABS and REL (and NOA) and ensure that the CPU and GPU results are bit-for-bit identical.

\section{Summary and Conclusions}
\label{sec:conclusion}

This paper explores correctness in error-bounded lossy quantizers. We describe problems that affect the ability to guarantee specific error bounds. We show code examples of how we addressed these issues in the LC compression framework we are developing and demonstrate that our fixes do not affect throughput but degrade the compression ratio (5\% on average). We hope our solutions will be helpful to others who work on lossy compression and will result in increased availability of guaranteed-error-bounded lossy compressors.

\begin{acks} % These are the same grants from the lossy paper
This work has been supported in part by the Department of Energy, Office of Science under Award Number DE-SC0022223 as well as by an equipment donation from NVIDIA Corporation. We thank Sheng Di and Franck Cappello for their invaluable help with domain knowledge during the development of LC.
\end{acks}

\bibliographystyle{ACM-Reference-Format}
\bibliography{refs}

\end{document}